\documentclass[a4paper,12pt]{article}
\usepackage{amssymb,amsmath,amsfonts}

\begin{document}

\author{Izumi OJIMA\\{\small \textit{Research Institute for Mathematical Sciences}}\\{\small \textit{Kyoto University, Kyoto 606-8502, Japan}}}
\title{A Simple Derivation of Josephson Formulae in Superconductivity}

\date{}
\maketitle
\begin{abstract}
A simple and general derivation of Josephson formulae for the tunneling
currents is presented on the basis of Sewell's general formulation of
superconductivity in use of off-diagonal long range order (ODLRO).
\end{abstract}

\vskip5pt

\vskip8pt

\section{Introduction}
According to the pioneering paper \cite{Haag62} by Haag on the BCS theory of
superconductivity \cite{BardCoopSchr57}, the general meaning of Cooper pair
condensates \cite{Cooper} can be understood as a ``variable at infinity'' 
\cite{Hepp72} which is based on the cluster property valid in any 
thermodynamic \textit{pure phases} described by factor 
states\footnote{A \textit{factor} state means a state such that the (GNS) 
representation of the algebra of physical quantities corresponding to it has a 
trivial centre.}. Along this line, Sewell \cite{Sewe97} takes, as a general 
characterization of the superconducting BCS states $\langle\cdot\rangle_{BCS}$ 
with condensed Cooper pairs, the off-diagonal long range order (ODLRO):
\begin{align}
& |\langle\psi(X_{1}+\frac{\xi_{1}}{2})\psi(X_{1}-\frac{\xi_{1}}{2})\psi
^{\dag}(X_{2}-\frac{\xi_{2}}{2})\psi^{\dag}(X_{2}+\frac{\xi_{2}}{2}%
)\rangle_{BCS}\nonumber\\
& \qquad\qquad\qquad\qquad-\Psi(X_{1},\xi_{1})\Psi^{\ast}(X_{2},\xi
_{2})|\underset{|\vec{X}_{1}-\vec{X}_{2}|-\infty}{\rightarrow}0.\label{ODLRO}%
\end{align}
Here $\psi(x)$ denotes the second-quantized non-relativistic electron field
obeying the canonical anticommutation relation%

\begin{equation}
\{\psi(x),\ \psi^{\dag}(y)\}=\delta^{3}(x-y),\label{CAR}%
\end{equation}
and the so-called ``macroscopic wave function'' $\Psi(X,\xi)=\langle
\psi(X+\frac{\xi}{2})\psi(X-\frac{\xi}{2})\rangle_{BCS}$ of a Cooper pair
should be non-vanishing in the limit of spatial infinity $|\vec{X}%
|\rightarrow\infty$. According to this formulation, he succeeded in giving a
general proof of the validity of Meissner effect ($\vec{B}=0$ inside of
superconductor) \cite{Sewe97}.

Inspired by these attempts to understand the essential features of
superconductivity in a model-independent way, I try here to present a simple
and general derivation of Josephson formulae\footnote{While this issue is
treated in \cite{Sewe97}, the arguments there are not clear or explicit
enough.} \cite{Jose62} which describe the tunnelling currents (\textit{dc}
and/or \textit{ac}) caused by the phase differences between two
superconductors separated by a thin barrier of the insulator. In contrast to
the traditional derivations based on the tunneling Hamiltonian \cite{Jose62},
we see that they are just a simple and direct consequence of the above ODLRO
and of the fact that the energy contained in one side of the junction gives a
non-trivial response to the global gauge transformation caused by the presence
of phase difference.

\section{Simple Derivation of Josephson Formulae}
Although a completely model-independent approach is desirable, the present
discussion requires the postulate of the standard BCS Hamiltonian
\cite{BardCoopSchr57,Haag62} arising effectively from the electron-phonon
coupling:
\begin{align}
& H_{BCS}(\Lambda;w)=\int_{\Lambda}dx[\frac{1}{2m}\vec{\nabla}\psi^{\dag
}(x)\cdot\vec{\nabla}\psi(x)-\mu\psi^{\dag}(x)\psi(x)]\nonumber\\
& +\frac{1}{|\Lambda|}\int_{\Lambda}dx\int_{\Lambda}dy\int_{\Lambda}%
dz\int_{\Lambda}du\ w(x,y,z,u)\psi^{\dag}(x)\psi^{\dag}(y)\psi(u)\psi
(z),\label{Hamil}%
\end{align}
where $m$ is the mass of an electron, $\mu$ the chemical potential of
electrons in the bulk superconductor occupying a macroscopically extended
spatial region $\Lambda$ with volume $|\Lambda|$. To exhibit the essence, we
use here a simple idealized picture of weakly coupled superconductors placed
in two spatial regions $\Lambda_{1}$, $\Lambda_{2}$ ($\subset\Lambda$)
separated by a Josephson junction regarded ideally as a \textit{phase boundary
}$W\equiv\partial\Lambda_{1}=\partial\Lambda_{2}$. In view of the wide
applicability of BCS model (at least in non high $T_{c}$ cases), the possible
differences in the properties of superconductors in $\Lambda_{1}$ and in
$\Lambda_{2}$ are supposed to be absorbed in the different choices of the
potential functions $w_{\Lambda_{1}}(x,y,z,u)\equiv w(x,y,z,u)$ for
$x,y,z,u\in\Lambda_{1}$ and $w_{\Lambda_{2}}(x,y,z,u)\equiv w(x,y,z,u)$ for
$x,y,z,u\in\Lambda_{2}$. Aside from this freedom, we understand that the
dynamics of superconductors are described universally by the above $H_{BCS}$
and that the differences of the realized thermodynamic phases are all reduced
to those in the choice of states $\langle\cdot\rangle$.

Here a remark need be added on a subtle point: To give a precise meaning to
the term ``thermodynamic \textit{phases}'', one should consider the theory in
the \textit{thermodynamic limit} with volume tending to infinity. Namely, the
sizes of regions $\Lambda_{1}$, $\Lambda_{2}$ (of course, finite on the
macroscopic scale) should be treated as ``infinitely large'' according to the
scale of microscopic interactions, and the location of the junction should be
supposed to be at ``spatial infinity'' far away from (the centre of)
$\Lambda_{1}$ in this scale. Although such expressions as this may sound quite
naive and vague lacking in the mathematical rigour, it is possible to give a
mathematically precise meaning to it in a non-standard analytic framework as
will be discussed briefly in the next section.

The next essential ingredient is the very definition of the tunneling electric
current flowing through the barrier. While our system does not have a
\textit{locally conserved} electric current owing to the presence of non-local
coupling in Eq.(\ref{Hamil}), the conservation of electric current is still
meaningful in the following sense: We define the electric charge $Q(\Lambda)$
in a spatial region $\Lambda$ by
\begin{equation}
Q(\Lambda)=-|e|\int_{\Lambda}dx\ \psi^{\dag}(x)\psi(x),
\end{equation}
which is conserved in the sense of
\begin{equation}
\lbrack H_{BCS}(\Lambda;w),\ Q(\Lambda)]=0,\label{conserv}%
\end{equation}
and which generates the electric global $U(1)$-gauge transformation:
\begin{equation}
\lbrack iQ(\Lambda),\ \psi(x)]=i|e|\psi(x),\ \ [iQ(\Lambda),\ \psi^{\dag
}(x)]=-i|e|\psi^{\dag}(x),
\end{equation}
where $e=-|e|$ is the unit of electric charge. In the situation with
$\Lambda=\Lambda_{1}\cup\Lambda_{2}\ (\cup W)$, the tunneling current $J$
between the two regions $\Lambda_{1}$ and $\Lambda_{2}$ is defined by
\begin{equation}
J=\frac{d}{dt}\langle Q(\Lambda_{1})\rangle_{BCS}=\langle\lbrack
iH_{BCS}(\Lambda),\ Q(\Lambda_{1})]\rangle_{BCS}.\label{Def}%
\end{equation}
Note that $J$ cannot be non-vanishing without the presence of the region
$\Lambda_{2}$ outside of $\Lambda_{1}$ in view of the simple equality
Eq.(\ref{conserv}) applied to $\Lambda=\Lambda_{1}$. (In a sense
Eq.(\ref{Def}) can be viewed as the basis of heuristic expression
$dN/dt=\partial H/\partial\theta$\ discussed in the number-phase picture of
Ginzburg-Landau theory.)

In view of the local (anti-)commutativity following from Eq.(\ref{CAR}), the
right-hand side of (\ref{Def}) can be reduced to
\begin{equation}
J=\langle\lbrack iH_{12},\ Q(\Lambda_{1})]\rangle_{BCS}=-\langle\lbrack
iQ(\Lambda_{1}),\ H_{12}]\rangle_{BCS}.\label{current}%
\end{equation}
Here, $H_{12}$ appears in the following decomposition of $H_{BCS}(\Lambda;w) $
corresponding to that of the spatial region $\Lambda$ into $\Lambda_{1}$ and
$\Lambda_{2}\ $(separated by $W$):
\begin{equation}
H_{BCS}(\Lambda;w)=H_{BCS}(\Lambda_{1};\frac{|\Lambda_{1}|}{|\Lambda
|}w)+H_{BCS}(\Lambda_{2};\frac{|\Lambda_{2}|}{|\Lambda|}w)+H_{12},
\end{equation}
according to which it can be written explicitly as
\begin{equation}
H_{12}=\frac{1}{|\Lambda|}\underset{\{i,j,k,l\}=\{1,2\}}{\sum}\int
_{\Lambda_{i}}dx\int_{\Lambda_{j}}dy\int_{\Lambda_{k}}dz\int_{\Lambda_{l}%
}du\ w(x,y,z,u)\psi^{\dag}(x)\psi^{\dag}(y)\psi(u)\psi(z).\label{Interaction}%
\end{equation}
Then the commutator in Eq.(\ref{current}) is calculated as
\begin{align}
& -\langle\lbrack iQ(\Lambda_{1}),\ H_{12}]\rangle_{BCS}=i|e|\frac{1}%
{|\Lambda|}\underset{\{i,j,k,l\}=\{1,2\}}{\sum}\int_{\Lambda_{i}}%
dx\int_{\Lambda_{j}}dy\int_{\Lambda_{k}}dz\int_{\Lambda_{l}}du\ \nonumber\\
& \times w(x,y,z,u)(-\delta_{i1}-\delta_{j1}+\delta_{k1}+\delta_{l1}%
)\langle\psi^{\dag}(x)\psi^{\dag}(y)\psi(u)\psi(z)\rangle_{BCS}.
\end{align}

To give the precise meaning of the ``macroscopic wave function'' $\Psi(X,\xi)$
in ODLRO (\ref{ODLRO}), we should now consider the thermodynamic limit of
$\Lambda,\Lambda_{1},\Lambda_{2}\longrightarrow\infty$, in which we assume the
ratio $|\Lambda_{1}|/|\Lambda_{2}|$ is kept fixed. When we evaluate the
right-hand side of (\ref{current}) in this situation, the contributions of
terms coming from the region $\Lambda_{2}$ can be replaced by their
expectation value in the BCS state $\langle\cdot\rangle_{BCS}$ because of the
cluster property justified by the understanding that the boundary $W$ is
infinitely far away in this limit, $\Lambda_{1},\Lambda_{2}\longrightarrow
\infty$ with $|\Lambda_{1}|/|\Lambda_{2}|$ kept fixed:
\begin{align}
& |\langle\psi^{\#}(x_{1}^{(1)})\cdots\psi^{\#}(x_{k_{1}}^{(1)})\psi
^{\#}(x_{1}^{(2)})\cdots\psi^{\#}(x_{k_{2}}^{(2)})\rangle_{BCS}\nonumber\\
& -\langle\psi^{\#}(x_{1}^{(1)})\cdots\psi^{\#}(x_{k_{1}}^{(1)})\rangle
_{BCS}\langle\psi^{\#}(x_{1}^{(2)})\cdots\psi^{\#}(x_{k_{2}}^{(2)}%
)\rangle_{BCS}|\nonumber\\
& \;\underset{\Lambda_{1},\Lambda_{2}\longrightarrow\infty,\ |\Lambda
_{1}|/|\Lambda_{2}|:fixed}{\longrightarrow}0,\label{Cluster}%
\end{align}
where $x_{1}^{(1)},\cdots,x_{k_{1}}^{(1)}\in\Lambda_{1},$ $x_{1}^{(2)}%
,\cdots,x_{k_{2}}^{(2)}\in\Lambda_{2}$ and $\psi^{\#}=\psi$ or $\psi^{\dag}$.
Since the expectation values of odd powers of fermionic operators $\psi,$
$\psi^{\dagger}$ vanishes\footnote{This holds under the moderate assumption of
spatial homogeneity at infinity \cite{BuchDopLonRob92}}, the only
contributions surviving this limit to the above integrand $\chi_{\Lambda_{i}%
}(x)\chi_{\Lambda_{j}}(y)\chi_{\Lambda_{k}}(z)\chi_{\Lambda_{l}}%
(u)(-\delta_{i1}-\delta_{j1}+\delta_{k1}+\delta_{l1})\langle\psi^{\dag}%
(x)\psi^{\dag}(y)\psi(u)\psi(z)\rangle_{BCS}\ $ come from the cases with
$i=j=1,k=l=2$ or $i=j=2,k=l=1$:
\begin{align}
& \underset{\{i,j,k,l\}=\{1,2\}}{\sum}\chi(x,y,z,u)(-\delta_{i1}-\delta
_{j1}+\delta_{k1}+\delta_{l1})\langle\psi^{\dag}(x)\psi^{\dag}(y)\psi
(u)\psi(z)\rangle_{BCS}\ \nonumber\\
& \underset{\Lambda_{1},\Lambda_{2}\longrightarrow\infty,\ |\Lambda
_{1}|/|\Lambda_{2}|:fixed}{\longrightarrow}\ \ \ \underset{i=j\neq k=l}{\sum
}\chi(x,y,z,u)(-\delta_{i1}-\delta_{j1}+\delta_{k1}+\delta_{l1})\nonumber\\
& \qquad\qquad\qquad\qquad\qquad\qquad\times\langle\psi^{\dag}(x)\psi^{\dag
}(y)\rangle_{BCS}\langle\psi(u)\psi(z)\rangle_{BCS}\ ,\label{factorize}%
\end{align}
where $\chi(x,y,z,u)\equiv\chi_{\Lambda_{i}}(x)$ $\chi_{\Lambda_{j}}%
(y)\chi_{\Lambda_{k}}(z)\chi_{\Lambda_{l}}(u)$ with $\chi_{\Lambda}$ being the
indicator function of a set $\Lambda$. If we assume the \textit{almost}
spatial homogeneity\footnote{This can be formulated without difficulty in the
non-standard analytic framework mentioned in Sect.3 in such a way that
suitable combinations of quantities involving the deviations from the
homogeneity are infinitesimal.} in each of regions $\Lambda_{1},\Lambda_{2}$
and if the potential $w(x,y,z,u)$ between electron pairs can be assumed to be
a real quantity, we can put
\begin{equation}
\langle\psi(x)\psi(y)\rangle_{BCS}\simeq\left\{
\begin{array}
[c]{l}%
|\Psi_{1}|e^{i\theta_{1}}\;(x,y\in\Lambda_{1}),\\
|\Psi_{2}|e^{i\theta_{2}}\;(x,y\in\Lambda_{2}),
\end{array}
\right. \label{CooperPhase}%
\end{equation}
from which we obtain the desired result for the $dc$-Josephson current by
picking out the phase factors from the integrands:
\begin{equation}
J\simeq\text{constant}\times(e^{i(\theta_{1}-\theta_{2})}-e^{-i(\theta
_{1}-\theta_{2})})\propto\sin(\Delta\theta),\label{Josephsn}%
\end{equation}
where $\Delta\theta\equiv\theta_{1}-\theta_{2}$ is the phase difference of
Cooper pairs across the junction.

Once this is obtained, it is straightforward to derive the formula for
$ac$-Josephson current applied to the situation with voltage gap $V$ across
the junction by replacing the above $\Delta\theta$ with $\Delta\theta+2eVt$.
While the postulated BCS Hamiltonian (\ref{Hamil}) with non-local interactions
is not fully compatible with a local gauge invariant coupling of the system
with spatially varying external electromagnetic field, it still allows a
coupling with $V$ in a way invariant under local gauge transformations with
only temporal dependence. As argued in \cite{Sewe97}, this gauge freedom
allows us to treat the situation with voltage gap $V$ simply by performing a
time-dependent local gauge transformation:
\begin{align}
A^{\mu} =(\phi,\ \vec{A}=0) & \longrightarrow (\phi+\frac{\partial}{\partial
t}(Vt)=V,\ \vec{A}=0);\\
\psi(x,t)  & \longrightarrow e^{ieVt}\psi(x,t),
\end{align}
which causes the above shift $\Delta\theta$ $\longrightarrow$ $\Delta
\theta+2eVt$ of the phase difference. Thus we have the formula $J_{ac}%
\propto\sin(\Delta\theta+2eVt)$ for $ac$-Josephson current.

In deriving Eq.(\ref{Josephsn}) we need the spatial homogeneity to extract the
factor of phase difference. Although the above kind of qualitative discussion
does not allow us to determine the precise coefficient, we can extract the
contribution to the energy density of the phase difference at the boundary
located infinitely far away in the similar way to the above, which gives us%

\begin{equation}
\frac{\langle H_{12}\rangle_{BCS}}{|\Lambda|}\underset{\Lambda_{1},\Lambda
_{2}\longrightarrow\infty,\ |\Lambda_{1}|/|\Lambda_{2}|:fixed}{\longrightarrow
}\text{constant}\times\cos(\Delta\theta).\label{Energy}%
\end{equation}
If the coefficient of $\cos(\Delta\theta)$ is assured to be of negative sign,
this guarantees the self-consistency of the postulate that the phase of Cooper
pair condensates in a superconductor should be spatially homogeneous in favour
of $\Delta\theta=0$ in the absence of such a constraint to maintain the phase
difference as the barrier. To verify this consistency problem in a more
satisfactory way, it seems necessary to confront a challenging problem of how
one can justify the notion of a \textit{point-like order parameter} $\Psi(x)$
of Cooper pairs, which appears in the Ginzburg-Landau phenomenological
approach \cite{GinzLand50}, and which is crucial for discussing the Type II
superconductivity involving in an essential way the spatial inhomogeneity and
the local gauge invariance problem.\footnote{This sort of situations may be
the very places where the conceptual advantages of non-standard analytic
formulation should be exhibited.}

\section{Discussion}
Although we refrain from a systematic explanation, we comment here how the
method of non-standard analysis \cite{StroLuxe78} can be useful in describing
the situation with infinitely large regions $\Lambda_{1}$, $\Lambda_{2}$
separated by a boundary $W$ at infinity. What is important is that it allows
us to treat both the finite volume theory and the infinite volume one
simultaneously \textit{without disconnecting} the two approaches. Before
introducing the distinctions among finite, infinity and infinitesimal (at the
syntactic level of \textit{internal objects}), everything looks as if we were
in the finite volume theory, but, once such distinctions are introduced (by
the interpretation in a \textit{non-standard model}) by regarding
$|\Lambda_{1}|$ and $|\Lambda_{2}|$ as \textit{infinite numbers} (whose ratio
is kept finite), the infinite volume theory is seen to be contained in the
former, being extracted through the procedure of taking \textit{finite parts}
of quantities which throws away all the infinitesimals such as $1/|\Lambda
_{1}|.$ In the situations with only one thermodynamic phase, this kind of
treament does not make much difference from the usual one, because the
procedure of extracting finite parts is more or less equivalent to taking the
thermodynamic limit. In the present situation with two infinitely large
regions $\Lambda_{1}$ and $\Lambda_{2}$, however, we have still ``another
world'' in $\Lambda_{2}$ \textit{beyond the infinitely distant} boundary $W$
of infinitely extended $\Lambda_{1}$. Whereas this situation seems difficult
to be accommodated in the usual formulation, it can be described without
difficulty in the framework mentioned here, where all the infinities and
infinitesimals are fully legitimate quantities. Moreover, all such limiting or
approximate relations as Eqs.(\ref{ODLRO}), (\ref{Cluster}), (\ref{factorize}%
), (\ref{CooperPhase}), (\ref{Josephsn}), (\ref{Energy}) are replaced by
simple algebraic equivalence relations modulo infinitesimals, in which one of
its conceptual advantages can be found.

\section*{Acknowledgments}
I am very grateful to Prof.~D.~Buchholz and Prof.~H.~Roos at Institut f\"ur
Theoretische Physik der Universit\"at G\"ottingen for critical comments on the
earlier version and for their warm hospitality extended to me in the summer of
1999. I thank Prof.~G.~Sewell very much for his kindness in sending me his two
papers in \cite{Sewe97} which were not accessible for me. Thanks are due to
Mr.~Hata for discussions. I have been partially supported by JSPS
Grants-in-Aid (No.11640113).

\end{document}